\documentclass[aps,prl,twocolumn,showpacs,groupedaddress]{revtex4}

\usepackage{graphicx}  
\usepackage{dcolumn}   
\usepackage{bm}        
\usepackage{amssymb}   
\usepackage{amsmath}
\usepackage{float}
\usepackage{color}
\usepackage{caption}

\begin{document}

\title{Bragg gravity-gradiometer using the $^1$S$_0$-$^3$P$_1$ intercombination transition of $^{88}$Sr}

\author{R P del Aguila$^{1}$, T Mazzoni$^{2}$, L Hu$^{1,3}$, L Salvi$^{1}$, G M Tino$^{1,4}$ and N Poli$^{1,5}$}
\address{$^{1}$Dipartimento di Fisica e Astronomia and LENS -
Universit\`{a} di Firenze, INFN - Sezione di Firenze, Via Sansone
1, 50019 Sesto Fiorentino, Italy \\  
$^{2}$now at LNE-SYRTE, Observatoire de Paris, 61 Avenue de l'Observatoire,
75014 Paris, France\\ $^{3}$ also at ICTP, Str. Costiera, 11 - 34151 Trieste, Italy \\ $^{4}$also at CNR-IFAC, Via Madonna del Piano, 10 - 50019 Sesto Fiorentino, Italy \\
$^{5}$also at CNR-INO, Via N. Carrara, 1 - 50019 Sesto Fiorentino, Italy}

\email{poli@lens.unifi.it}

\date{\today}

\begin{abstract} We present a gradiometer based on matter-wave
interference of alkaline-earth-metal atoms, namely $^{88}$Sr. The
coherent manipulation of the atomic external degrees of freedom
is obtained by large-momentum-transfer Bragg diffraction, driven
by laser fields detuned away from the narrow $^1$S$_0$-$^3$P$_1$ intercombination
transition. We use a well-controlled artificial gradient,
realized by changing the relative frequencies of the Bragg
pulses during the interferometer sequence, in order to characterize the sensitivity of the gradiometer.  The sensitivity reaches $1.5 \times 10^{-5}$ s$^{-2}$ for an interferometer time of 20 ms, limited only by geometrical constraints. 
We observed extremely low sensitivity of the gradiometric phase to magnetic field gradients, approaching a value 10$^{5}$ times lower than the sensitivity of alkali-atom based gradiometers. An efficient double-launch technique employing accelerated red vertical lattices from a single magneto-optical trap cloud is also demonstrated.  These results highlight strontium as an ideal candidate for precision measurements of gravity gradients, with potential application in future precision tests of fundamental physics.
\end{abstract}
		
\maketitle

\section{Introduction}

Matter-wave atom interferometry has rapidly grown in the last decade and is proving to be a powerful tool for investigation of fundamental and applied physics \cite{tino2014atom}. Precision interferometric devices are of particular interest in gravitational physics, where they allow precise measurements of gravity acceleration \cite{Peters1999}, gravity gradients \cite{Sorrentino2014, Asenbaum2016}, gravity curvatures \cite{Rosi2015}  and the Newtonian gravitational constant \cite{Rosi2014}. The investigation of novel interferometric schemes which implement atomic species other than the more commonly used alkali species is seeing increasing demand, particularly for dramatic improvements of fundamental tests of General Relativity \cite{Tarallo2014, Rosi2017, Biedermann2015,Zhou2015,Duan2016} and gravitational wave detection in the low-frequency regime \cite{Dimopoulos2009,Hohensee2011,Graham2013}. Increasing the precision and sensitivity of interferometric metrology devices, as well as understanding and characterizing the limitations of novel interferometric schemes with non-alkali atoms is an important step towards the goal of heralding a new generation of viable precision measurement devices to be employed in the search of new physics \cite{Safronova2017}.


In this article, we demonstrate the first differential two-photons Bragg interferometer based on the intercombination transition of strontium atoms. This forbidden transition is a thousand times narrower than the transitions previously employed in experiments carried on with alkali and alkali-earth atoms. Moreover, taking advantage of particular properties of $^{88}$Sr isotope, we demonstrate a high-contrast gradiometer with an extremely low sensitivity to magnetic field gradients, about one hundred thousand times less sensitive than alkali-atom based gradiometers.

\section{Background}\label{sec.background}
The interest in alkaline-earth-metal (-like) atoms for precision interferometry has grown rapidly during the last decade because of their unique characteristics \cite{Tarallo2014,Graham2013,Riehle1991,Jamison2014,Hartwig2015,Mazzoni2015,Zhang2016}. For instance, their $^1$S$_0$ ground state has zero angular momentum and, in particular, bosonic atoms such as the $^{88}$Sr isotope do not even have a nuclear spin, so their ground state has zero magnetic moment at first order. This leads to ground-state $^{88}$Sr being extremely insensitive to stray magnetic fields, about five orders of magnitude less sensitive than alkali atoms. Another characteristic of alkali-earth-like atoms is their two-valence-electron structure, which leads to the presence of narrow intercombination transitions. For $^{88}$Sr, the $^1$S$_0$-$^3$P$_1$ triplet transition has a highly favorable $\sim7$ kHz linewidth. This transition can be used for efficient Doppler laser cooling down to the recoil temperature and it has recently been employed for the fast production of degenerate gases of strontium atoms \cite{Stellmer2013}. 
Moreover,  ground-state $^{88}$Sr has a uniquely negligible s-wave scattering length of $a=-2a_{0}$, which makes this atom very insensitive to cold collisions. Thanks to this feature, Bloch oscillations of ultra-cold $^{88}$Sr atoms trapped in vertical optical lattices were observed with long coherence times \cite{Poli2011}.

In pulsed atom interferometry the matter-wave interference is realized by splitting the atomic wave packet in a coherent superposition of two states (internal and/or external) and recombining them after a free-evolution time $T$, by means of standing-wave pulses, namely Raman or Bragg transitions. Because of the absence of a hyperfine structure in the ground state, Raman transitions are not available for $^{88}$Sr; instead, Bragg diffraction can still be employed to coherently control the atomic momentum. Bragg diffractions have the advantage of keeping the atom in the same internal (electronic) state, so multiple pairs of photons can be exchanged between the optical standing-wave and the atom in a single interaction \cite{Martin1988,Giltner1995b}. Thanks to this mechanism, large-momentum-transfer schemes can be realized in pulsed atom interferometers \cite{Muller2008b}. The momentum splitting given by an $n$-order Bragg transition is $\hbar k_\mathrm{eff} = 2n\hbar k$, where $k=2\pi/\lambda$ is the wave vector of the Bragg laser with wavelength $\lambda$. Large-momentum-transfer schemes allow the interferometer to have an increased sensitivity to phase shifts \cite{Mazzoni2015}. Since the atom remains in the same electronic state during a Bragg transition, systematic effects such as light shift are suppressed \cite{Altin2013}.

In contrast to previous experiments in which we have driven Bragg transitions with laser beams detuned away from the strong $^1$S$_0$-$^1$P$_1$ ``blue" transition at 461~nm \cite{Mazzoni2015,Zhang2016}, in this work we have used 689~nm ``red" light which is detuned away from the $^1$S$_0$-$^3$P$_1$ intercombination transition. 

The particular combination of the much smaller linewidth of this transition ($\Gamma_R=2\pi\times 7.6$~kHz = $2\times10^{-4} \Gamma_B$ in units of the linewidth of the dipole allowed blue transition $\Gamma_B$) and the much higher available laser power at 689~nm makes this transition particularly favorable for Bragg diffraction. 

Indeed, at equal laser intensities and for equal two-photons Rabi frequencies, the estimated scattering rate in a Bragg diffraction process depends only on the Bragg order $n$. In particular, for $n=2$ the single photon scattering rate in the red is four times less than the scattering rate calculated in the blue. 
Furthermore, the higher laser power available at red wavelengths allows operation at a much larger relative detuning from resonance than when working with the blue transition ($\Delta/\Gamma_R > 10^2 \Delta/\Gamma_B$) while keeping similar Rabi frequencies.

As a result of these facts, there are several benefits of atom interferometers performed on the narrow intercombination transition of $^{88}$Sr atoms as presented in the following sections. In particular: the much higher interferometer contrast than previously obtained with the blue transition and the possibility to employ the same red light for efficient double-launches from a single magneto-optical trap, through fast frequency tuning of the trapping red light across the narrow transition. Indeed, this configuration represents a great simplification over previous gradiometer and gravimeter launch sequences realized with strontium atoms \cite{Mazzoni2015, Zhang2016}. Furthermore, we demonstrated for the first time the expected ultra-low sensitivity to magnetic field gradients of a strontium  atomic gradiometer.


\begin{figure*}
	\includegraphics[width=1.0\linewidth]{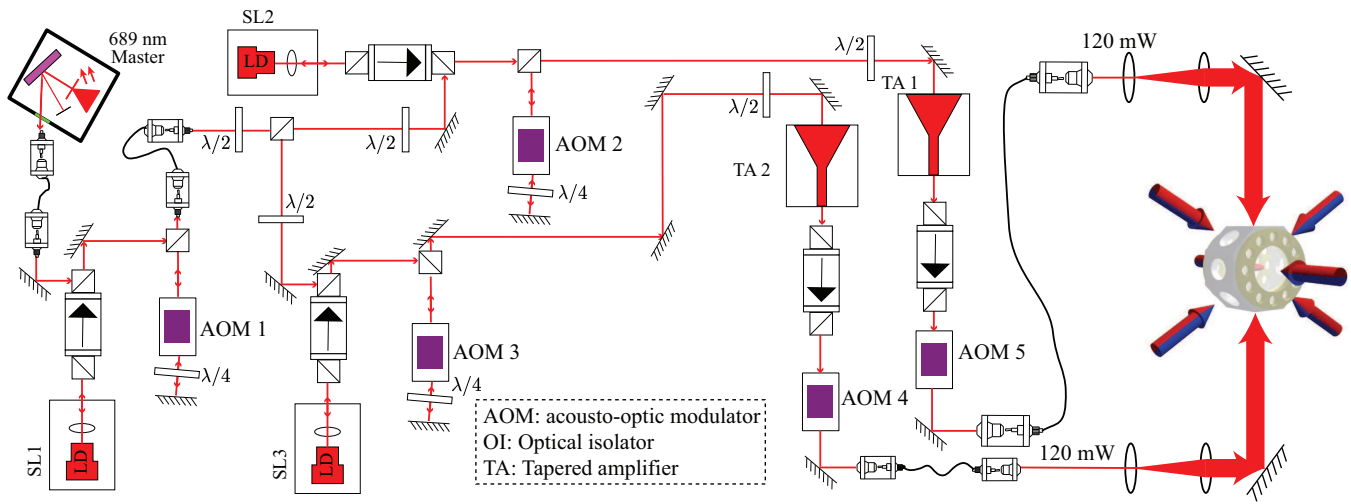}
	\caption{Schematic representation of the red Bragg laser setup. The source is based on a sub-kHz master laser at 689~nm, referenced to the $^{88}$Sr intercombination transition, which injection-locks a series of slave diode laser (SL1, SL2, SL3) amplified with two tapered amplifiers (TA1, TA2). A first acousto-optic modulator (AOM1) is employed to set the detuning from resonance for both Bragg beams. The light coming out from this AOM is then sent to two secondary slave lasers, SL2 and SL3, which produce the two Bragg beams. Their relative frequencies are set by AOM2 and AOM3 in order to create standing or traveling waves and to match the Bragg resonance condition. Light beams coming out these two AOMs are power amplified through two independent tapered amplifiers (TA1 \& TA2). A final set of AOMs (AOM4 \& AOM5) is employed to shape the Bragg pulses with a Gaussian amplitude profile. The two beams are then injected into polarization-maintaining fibers and sent to the atomic sample vertically along opposite directions.} \label{fig.LaserScheme}
\end{figure*}

\section{Experimental setup and methods} \label{sec.expsetup} 
The experimental setup for cooling and trapping $^{88}$Sr atoms is similar to the setup used in earlier Bragg interferometry experiments, previously reported in \cite{Mazzoni2015,Zhang2016}. The main difference consists of a new laser scheme based on red lasers tuned at 689~nm, adopted to create the traveling and standing waves (Bragg pulses, optical lattice trapping) necessary to manipulate the atomic momentum (see figure~\ref{fig.LaserScheme}). In brief, it relies on an optically-amplified sub-kHz linewidth laser source at 689~nm composed of a master laser (frequency stabilized external-cavity diode laser, referenced to the intercombination transition \cite{Poli2006}) and a set of slave diode lasers/tapered amplifiers. A first slave diode laser (SL1), injection-locked to the master, is used to set the main detuning $\Delta$ of the Bragg pulses from the atomic resonance. With the use of a double-pass acousto-optical modulator (AOM1) it is possible to change the detuning in the range $-95$~MHz$<\Delta<+145$~MHz ($-1.2\times10^4< \Delta/\Gamma_R < +1.9\times10^4$). The two Bragg beams are generated by two independent tapered amplifiers, seeded by two separate slave lasers (SL2, SL3) optically injected by SL1. The relative frequency between the two Bragg beams is set by two independent double-pass AOMs (AOM2 \& AOM3) to match the Bragg resonance condition for the free-falling atoms and to generate the accelerated lattice, to initially launch upward the atoms. Frequency ramps for the AOMs are generated by programmable direct digital synthesizers (DDS). The two beams are independently shaped in amplitude (two additional AOMs provide a Gaussian amplitude profile \cite{Muller2008}) and sent to the atoms via polarization maintaining (PM) fibers. The power available at each fiber output is about $120~$mW. Both beams are shaped and collimated to a $1/e^2$ radius of $w_0=2.25$~mm.

The experimental sequence is as follows. Ultra-cold $^{88}$Sr sample is produced in a two-stage magneto-optical trap (MOT), as described previously \cite{Mazzoni2015,Zhang2016}. About $2\times10^6$  atoms are trapped in $1.5$~s, with a temperature of 1.2~$\mu$K and a spatial radial (vertical) size of $300~\mu$m ($50 ~\mu$m) full-width half-maximum (FWHM). After the MOT is released, about 50\% of the atoms are adiabatically loaded over 100~$\mu$s into an optical lattice, realized by the two counter-propagating red Bragg laser beams. With a detuning $\Delta=-95$~MHz, the lattice trap depth is $U=20E_r$ in recoil units (where $E_r=(\hbar k)^2/2m$ is the recoil energy of $^{88}$Sr atoms for 689 nm photons). The atoms remain in the stationary lattice for about $500~\mu$s to allow the magnetic fields from the MOT stage to fully dissipate, after which they are accelerated upwards at a rate of $30~g$ (where $g$ is acceleration due to gravity) in about 3~ms, by frequency chirping the upper red beam. The launched atoms are then adiabatically released from the accelerated lattice in $90~\mu$s.

After a time $T_s$ (typically $10$~ms\,\,\,$<T_s<30$~ms corresponding to a gradiometer baseline $2.7$~cm\,\,\,$<\Delta z<3.9$~cm) the same launch procedure is repeated by trapping the residual free-falling atoms from the MOT. In this case, by adjusting the second launch duration, it is then possible to precisely set the relative final velocities of the two launched clouds. This procedure also ensures that the final launch frequency for the second launch is lower than that of the first launch, preventing interactions between the first launched cloud and the second accelerating lattice. For the gradiometer, we set the launch parameters to produce two clouds of $5\times10^5$ atoms each, with a center-of-mass momentum difference of 36~$\hbar k$ (where $v_r=\hbar k/m=6.6$~mm/s is the recoil velocity for 689~nm photons).

\begin{figure*}
	\includegraphics[width=1.0\textwidth]{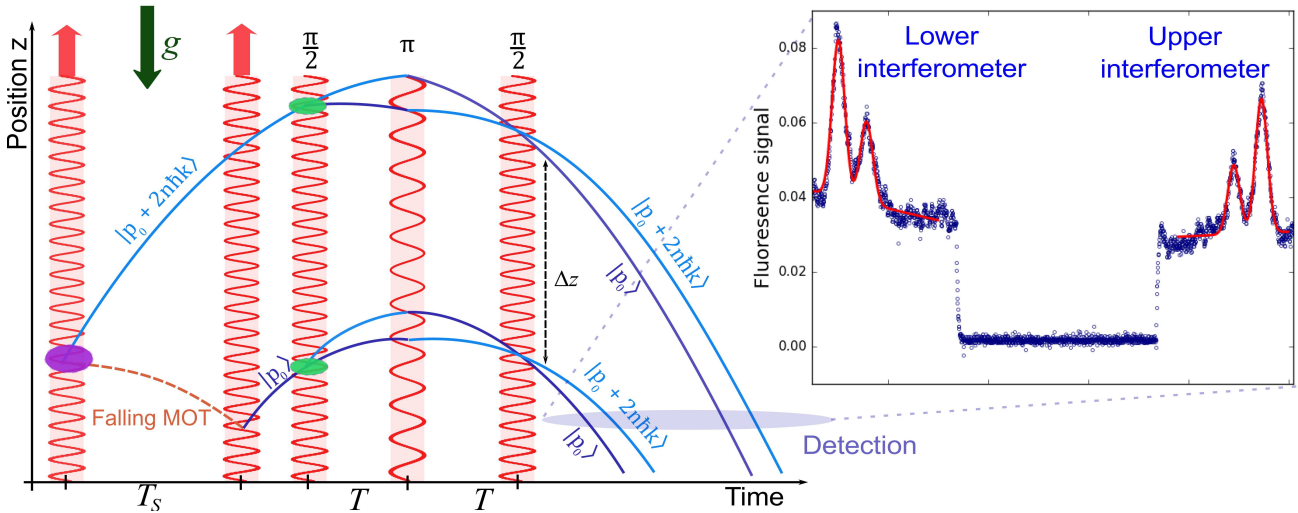}
	\caption{ Bragg gravity gradiometer experimental sequence. Two $^{88}$Sr atomic clouds are launched upwards (z-axis) in a fountain with accelerated ($30~g$) optical lattices. The separation time $T_s$ between the launches sets the baseline $\Delta z$ for the gradiometer. After velocity selection, the lower and upper clouds are in a momentum state $|p_0\rangle$ and $|p_0 + 2n \hbar k_\mathrm{blue}$ respectively. A sequence of $\pi/2-\pi-\pi/2$ Bragg pulses interacts simultaneously with both clouds generating a Mach-Zehnder interferometer with momentum splitting $\pm 2n\hbar k\mathrm{blue}$ and pulse spacing $T$. The inset shows a typical detection signal, with the two arms of each interferometer resolved using fluorescence detection. A Gaussian fit of the signal peaks resolves the number of atoms in each momentum state, giving the relative population for each interferometer.} \label{fig.Fig2}
\end{figure*}

After the launch stage, the two clouds are each velocity-selected by an individual sequence of Bragg $\pi$-pulses in order to narrow the momentum spread before the interferometer sequence. An initial {35~$\mu$s}-long 1$^\textrm{st}$-order ($n=1$) pulse selects a narrow momentum distribution, and a following set of {25~$\mu$s}-long 2$^\textrm{nd}$-order ($n=2$) pulses spatially separates the selected cloud from the residual launched cloud. The sequence for each cloud differs in the total number of pulses and in the direction of momentum imparted. This results in two velocity-selected clouds of about $5\times10^4$ atoms with a momentum spread of 0.15~$\hbar k$, separated in momentum by precisely $\Delta p=4\hbar k$. This guarantees that both clouds will interact simultaneously with all the 2$^\textrm{nd}$-order Bragg pulses we use for the interferometer. The entire launch and selection stages take 50~ms. The Mach-Zehnder-like interferometer sequence consists of three {25~$\mu$s}-long 2$^\textrm{nd}$-order Bragg pulses, equally separated by a time $T$ (up to 25~ms). In order to get the exact mirror and beam-splitter pulses, the amplitude of each pulse is properly tuned.

Finally, the two output ports of the two simultaneous interferometers are detected in time-of-flight by collecting the fluorescence signal induced on the dipole allowed transition.  The detection is done about 40~ms after the last pulse is applied (see inset in figure~\ref{fig.Fig1}), when the two momentum states of each interferometer are sufficiently separated in space. The population at each output port is then determined through Gaussian fits of the respective fluorescence signal. The relative population for each interferometer is plotted one against the other, in order to obtain an ellipse, from which the relative phase can be extracted \cite{Foster2002}.

\subsection{Artificial gradient generation}\label{sec.ArtificialGradient}
In order to characterize the sensitivity of our gradiometer to relative phase shifts, we induced a well-controlled artificial gradient between the two interferometers, during the interferometer sequence. The method initially proposed to compensate for the loss of contrast due to gravity gradients in atom interferometers \cite{Roura2015}, is based on the use of interferometer pulses with differing effective wavevector $k_\mathrm{eff}$. Specifically, an artificial $\Gamma_\mathrm{artif}$ is realized by changing the relative wavelength between the beam-splitter pulses ($\pi/2$-pulse with wavevector $k_\mathrm{eff}$) and the mirror pulse ($\pi$-pulse with wavevector  $k_\mathrm{eff}+\Delta k_\mathrm{eff}$). The main effect of this change is to unbalance the momentum transfer between the two branches of the interferometer, generating an additional phase shift term, which depends on the initial position and velocity of the atoms \cite{Roura2015}. In the gradiometer configuration, we expect an additional phase shift term as:
\begin{equation}
    \label{eq.artifphase}
    \Delta\phi_\mathrm{artif} = -2 \Delta k_\mathrm{eff} (\Delta z + \Delta v T).
\end{equation}

This extra term can be interpreted as an artificial gradient along the vertical $z$ direction with an amplitude $\Gamma_\mathrm{artif} = 2\Delta k_\mathrm{eff}/k_\mathrm{eff}\,T^2$. In our experiment, we are able to control $k_{\mathrm{eff}}$ through the use of AOM1 (by applying a frequency jump $\Delta_\pi$ between pulses), and $\Delta z$ by setting the time $T_s$ between the two successive launches, both with extremely high precision. By using this method it is then possible to set a specific phase offset between the two interferometers and a non-degenerate gradiometer ellipse graph, suited for the determination of gradiometer sensitivity. The total phase difference between the two arms of the gradiometer, when incorporating the artificial gradient is:
\begin{equation}
		\label{eq.gradiophase}
		\Delta\Phi = - k_{\mathrm{eff}} (\Gamma + \Gamma_\mathrm{artif}) T^2 (\Delta z + \Delta vT)
\end{equation}
where $\Gamma \sim -	3 \times 10^{-6}$~s~$^{-2}$ is the gradient due to Earth's gravity.

Compared to other magnetically-induced phase-shift methods \cite{Foster2002,Wang2016}, used to control ellipse phase and to characterize gradiometer sensitivity, the artificial gradient method relies only on optical frequency jumps, which can be controlled with much higher precision. Moreover, the possibility to drive Bragg pulses close to a narrow transition, allows the use of a single AOM to easily drive $\pi/2$- and $\pi$-pulses symmetrically displaced to the red and blue side of the resonance, maintaining identical Rabi frequency and scattering rates. 

\section{Experimental results} \label{sec.expresults}

	\subsection{Interferometer contrast}
\begin{figure}
    \begin{center}
        \includegraphics[width=0.5 \textwidth]{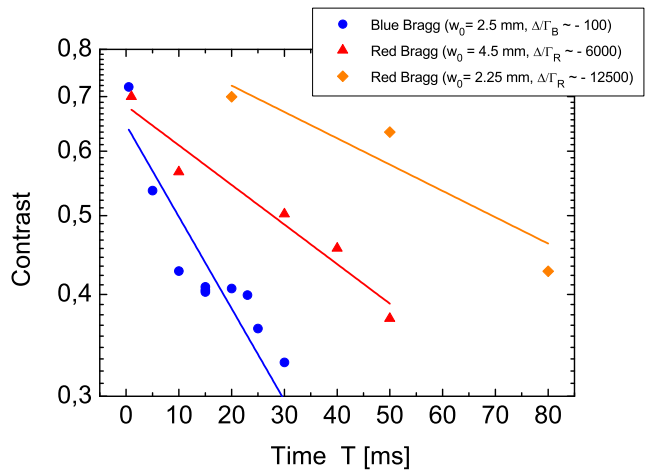}
        \caption{Comparison of Mach-Zehnder interferometer contrast with ``blue" and ``red" Bragg pulses, respectively driven by laser beams near the strong $^1$S$_0$-$^1$P$_1$ blue transition at 461~nm (blue circles) and the narrow $^1$S$_0$-$^3$P$_1$ red intercombination transition at 689~nm (red triangles and orange diamonds), for different beam radii $w_0$. The contrast obtained with red Bragg beams is always higher than with the blue ones. The contrast decay times from an exponential fit of the data (blue, red and orange lines) are $\tau_B=39(6)$~ms and $\tau_{R1}=89(13)$~ms, $\tau_{R2}=130(50)$~ms respectively for blue, and red Bragg. For the red Bragg, contrast and contrast decay are improved for smaller beam radii and larger detuning.}
        \label{fig.ContrastComparison}
    \end{center}
\end{figure}

	We investigated the benefit of driving Bragg transitions using the intercombination transition by comparing the observed contrast of single Mach-Zehnder interferometers. Under typical parameters, with $25$~$\mu$s-long Bragg interferometer pulses, the interferometer contrast as a function of the interferometer time $T$ is significantly higher than the contrast obtained with Bragg transitions on the strong blue transition (see figure~\ref{fig.ContrastComparison}). As anticipated, we attribute this result to the much lower single-photon scattering rate during each Bragg pulse on the intercombination transition, also thanks to the much larger relative detuning $\delta/\Gamma$. The result of this is not only higher contrast, reaching $C=0.42$ for $T=80$~ms, (obtained for a relative detuning $\Delta/\Gamma_R=1.25\times10^4$ and a Bragg beam radius of $2.25$~mm),  but also lower contrast decay rate. From an exponential fit of the data in figure~\ref{fig.ContrastComparison} we obtain a maximum contrast decay time of $\tau_{R2}=130(50)$~ms, which represents an improvement of a factor of about three, with respect to the contrast decay observed with blue Bragg ($\tau_{B}=39(6)$~ms). 
	It is important to notice that further improvements in the contrast decay rate are foreseen, by reducing the radial expansion of the atomic cloud and Bragg beams wavefront aberrations, as already suggested in previous work \cite{Mazzoni2015, Schkolnik2015}.		

\subsection{Lattice launch efficiency}

\begin{figure*}		
  \begin{center}
		\includegraphics[width=0.9\textwidth]{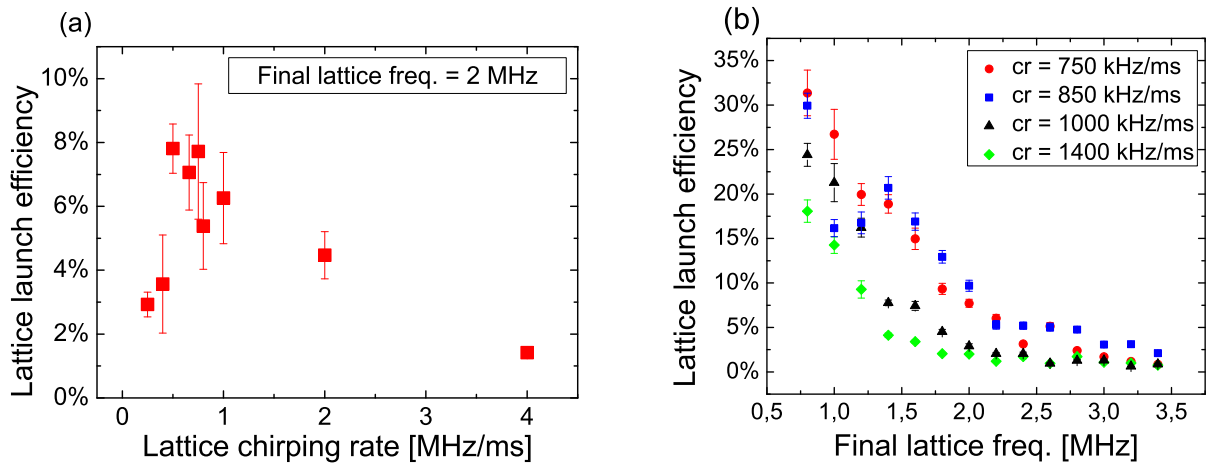}
    \caption{Lattice launch efficiency as a function of frequency chirp rate (a) and final lattice frequency detuning (b). For a fixed final lattice frequency detuning of 2~MHz the launch chirping rate that optimizes the launch efficiency ($\sim8\%$) is close to 850~kHz/ms. A larger launch efficiency can be obtained at the expense of final lattice detuning (e.g. final cloud vertical height). Typical final lattice detuning is set to 2.8~MHz (2.2~MHz) for the first (second) launch. These parameters guarantee a sufficient height of both clouds for the interferometer sequence and for separate fluorescence detection of momentum state in time of flight. In these condition about $5 \times 10^4$ atoms are launched in each cloud.}
       \label{fig.LaunchEff}
   \end{center}
\end{figure*}

	One of the advantages of the red laser system lies in the possibility to employ it for an efficient double-launch sequence with accelerated lattices. Compared to the more commonly used ``juggling'' technique \cite{Legere1998,Bertoldi2006,Duan2014}, in which the two gradiometer clouds are obtained with two separate MOTs, we can make more efficient use of the atoms prepared in a single red MOT, eventually resulting in a tremendous reduction of the total cycle time of the gradiometer.  We characterized the trapping and launch efficiency, in order to find the best launch parameters, by measuring the number of atoms available for the interferometer sequence at the end of the launch. Figure~\ref{fig.LaunchEff} shows the efficiency of the launch as a function of two different lattice parameters: the upper red lattice beam chirping rate (setting the lattice acceleration), and the final frequency detuning (setting the final velocity of the launched cloud).

The launch is typically performed by choosing an absolute detuning of both lattice beams from the resonance of about $\Delta=-95$~MHz. The choice of detuning was determined experimentally to maximize the number of atoms available after the launch. In general, we took care to work with detunings from resonance far from the photo-association line at $\Delta=-24$~MHz, both for the launch and the interferometer, since the change in the scattering length would eventually produce unwanted phase shifts in the gradiometer \cite{Blatt2011}. Given a total intensity of lattice beams on the atom of about $950$~mW/cm$^2$, we estimate a scattering rate in this condition of about $60$~s$^{-1}$ and a lattice trap depth of $U=20~E_r$.

An efficient launch requires fulfilling the condition for the acceleration to be lower than the critical acceleration $a_c$, to avoid Landau-Zener tunneling \cite{Peik1997}. In our condition, we estimate a maximum possible acceleration of $a_c=4\times10^3$~m/s$^2$.  Due to the finite lattice lifetime, the lattice launch is then performed typically over a short time, with considerably large accelerations. Setting a typical launch height to $2.5$~cm, corresponding to a final relative frequency detuning of 2~MHz, we found an optimum value for the lattice beam chirping rate of 850~kHz/ms, corresponding to an acceleration of  $30$~$g$. Under these conditions, we obtain comparable launch efficiencies (up to 10\%, see figure~\ref{fig.LaunchEff})) with respect to lattice launch efficiencies obtained with far-detuned lattice laser light, as previously reported in \cite{Zhang2016}. 

In a typical experimental cycle, the launch sequence is repeated two times with similar parameters. In terms of absolute atom number we typically obtain about $5\times10^4$ atoms in each cloud, enough to provide a sufficient signal at detection. 
Higher efficiencies have been observed (up to 32 \%, see figure~\ref{fig.LaunchEff}(b) for smaller final lattice frequencies ($750$~kHz) and for smaller lattice chirp rate ($750$~kHz/ms). This efficiency is a combined effect of a reduced chirp rate and a reduced launch time, which, in the latter configuration is only $1$~ms, indicating additional channels of fast atom losses. To explore this conjecture, we performed lifetime measurements of atoms held in a static red lattice. The observed lifetime for a steady $689$~nm lattice for $\Delta = -95$~MHz is only $\sim 30$~ms, a value almost $20$ times smaller than the expected value estimated by solely single-photon resonant scattering events.  This indicates clearly that additional mechanisms such as parametric heating effects \cite{Jaregui2001} and additional contributions to resonant scattering from the spontaneous emission spectrum of red tapered amplifiers are strongly limiting the lattice lifetime. We interpret these as the main limitations for the observed launch efficiency for launch times longer than few ms. Indeed, being only a technical limitation, we expect that the use of quieter lasers, with lower intensity noise and smaller spontaneous emission (for example by using solid-state Ti:Sa laser systems), would result in a large improvement in launch efficiency also for longer launch durations.

	\subsection{Relative phase shift sensitivity}

\begin{figure*}
	\begin{center}
		\includegraphics[width=1. \textwidth]{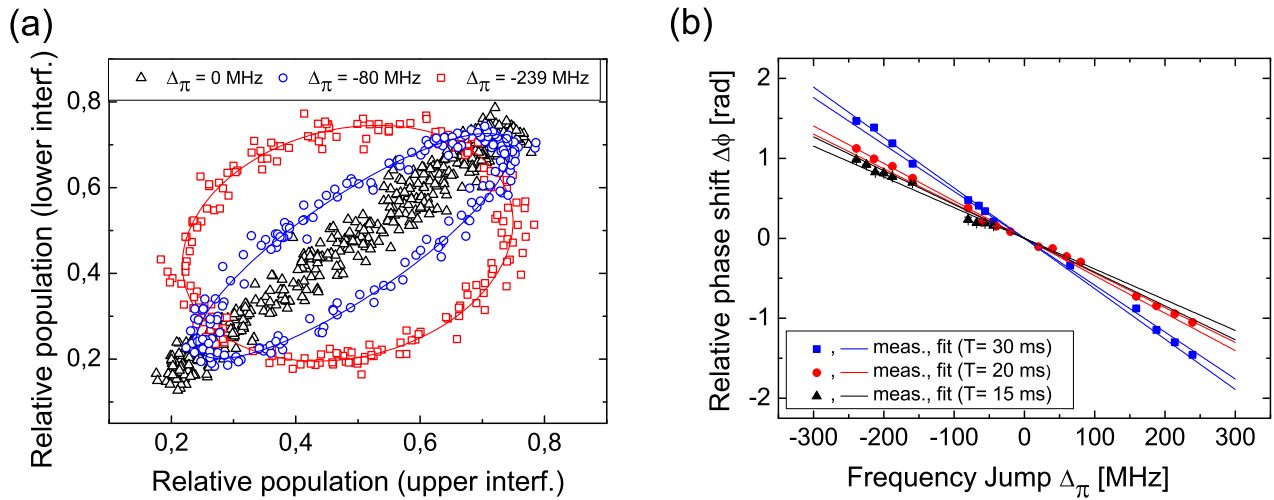}
		\caption{(a) Measured relative populations for the upper and lower interferometers
			(plotted one versus the other) for different $\pi$-pulse frequency
			jump $\Delta_\pi$ with gradiometer time $T=20$ ms. Changing the detuning for the $\pi$-pulse
			induces an artificial gradient; increasing the artificial
			gradient opens the ellipse progressively, demonstrating the
			appearance of a fixed relative phase between the upper and lower
			interferometers. (b) Relative phase shifts obtained by least-square ellipse
			fitting for different frequency jumps $\Delta_\pi$ at three
			different separations $\Delta z$: $\Delta z=2.7$~cm (black
			triangles, $T=15$~ms), $\Delta z=3.2$~cm (red circles, $T=20$~ms) and $\Delta z=3.6$~cm
			(blue squares, $T=30$~ms). Lines represent 95\% confidence levels of the fitted dataset. 
			Fitted values of the relative phase shift as a function of the frequency jump are consistent with the theoretical estimations from Eq. \ref{eq.artifphase} and Eq.~\ref{eq.gradiophase} within the confidence level.}\label{fig.ellipses}
	\end{center}
\end{figure*}

We characterized the sensitivity of our gradiometer by including a well-controlled artificial gradient between the two interferometers. Figure~\ref{fig.ellipses}(a) shows the obtained gradiometric ellipses for differing detuning jumps $\Delta_\pi$ between the $\pi/2$- and $\pi$-pulses, for $T=20$~ms. In our case, a relative shift of almost $\pi/2$ can be induced for our maximum detuning jump $\Delta_\pi=239$~MHz (red squares). This technique allows us to induce a large additional relative shift between the two clouds. In our case, the effect of gravity gradients is in fact too small with respect to the current sensitivity (black triangles).

Figure~\ref{fig.ellipses}(b) shows the measured relative phase shifts for two clouds with a velocity separation of $\Delta v=4\hbar k_r/m$ and three different cloud separations: $\Delta z=2.7$~cm (black squares), $\Delta z=3.2$~cm (red circles) and $\Delta z=3.6$~cm (blue triangles). In each case, the measured phase shift agrees with the expected phase estimated from Eq. \ref{eq.gradiophase}.

\begin{figure}[t]
	\includegraphics[width=0.45  \textwidth]{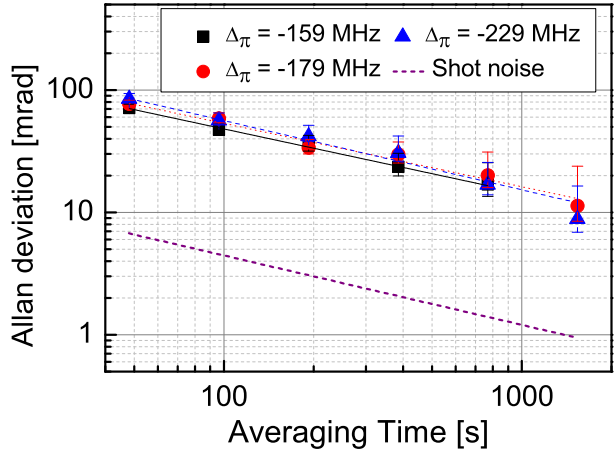}
	\caption{\indent Allan deviation of the relative phase shift for
		a dual interferometer with 2nd-order Bragg pulses, $T=20$~ms and a
		baseline $\Delta z=3.2$~cm with an artificial gradient
		corresponding to three different detuning frequency jumps. The Allan deviation
		was calculated with 20 points per ellipse, and it scales as $\tau^{-1/2}$ (red line) showing a
		sensitivity at 1~s of 210~mrad.} \label{fig.Allan}
\end{figure}

	We estimated the Allan deviation of the measured phase shift to measure the short-term sensitivity of our gradiometer. Figure~\ref{fig.Allan} shows the Allan deviation of three independent sets of
	3740 measurements each for $T=20$~ms and gradiometer baseline $\Delta z=3.2$~cm.  The relative phase shifts were obtained by inducing an artificial gradient of up to $\Gamma_\mathrm{artif}=2.8$~s$^{-2}	$ with a frequency jump respectively of  $\Delta_\pi=-159$~MHz (black squares), $\Delta_\pi=-179$~MHz (red circles) and $\Delta_\pi=-229$~MHz (blue triangles).  The cycle time was set to 2.4~s for an overall measurement time of about 2.5~h.

	For all the datasets, the Allan deviation scales as $\tau^{-1/2}$ (where $\tau$ is the averaging time) showing that the main noise contribution comes from white phase noise. The relative phase sensitivity at 1~s is 210~mrad which, for our experimental parameters (2nd-order Bragg pulses, $T=20$~ms, $\Delta z=2.7$~cm), corresponds to a sensitivity to gravity gradients of $5\times10^{-4}$~s$^{-2}$.

	Integrating up to 1000 s, we reached a best sensitivity to gravity gradients of $1.5\times10^{-5}$~s$^{-2}$, mainly limited by detection noise due to the limited optical access of our chamber. As a comparison, figure~\ref{fig.Allan} also shows the estimated shot-noise limit \cite{Sorrentino2014}, which lies about a factor of 10 below the current experimental sensitivity. It is worth noticing that limitations to the present level of sensitivity are mostly technical and not fundamental. Improvements in the current atom trapping and detection chamber are foreseen in order to increase the atom detection efficiency and the interferometer time $T$. Indeed, based on these results, no fundamental limitation is foreseen in reaching the state-of-the-art gravity gradiometry sensitivity of rubidium (Rb) atom interferometers.

\subsection{Magnetic field sensitivity}

Given the particular level structure of Sr atoms, with specific reference to zero-spin bosonic isotopes, it is expected that a Sr Bragg interferometer will be largely insensitive to magnetic fields. Considering only the effect coming from the magnetic moment of the $^1$S$_0$ ground state, the calculated theoretical value of the shift at the second order in the field amplitude comes from the diamagnetic term in the Hamiltonian and it is $\beta =$~	5.5~mHz/G$^{2}$ \cite{safronovaPrivate}. Following the argument presented in \cite{Damico2016}, in the presence of a magnetic field gradient $B'$, this term will produce a relative phase shift in the gradiometer:

\begin{equation}
		\label{eq.magneticphase}
		\Delta\phi_M = 4 \pi \frac{2 n \hbar k}{m}	\beta	 T^{2} (B_{0}^{u}B'^{u}-B_{0}^{l}B'^{l})
\end{equation}

where $B_0^{u(l)}$ is the static field magnitude and $B'^{u(l)}$ is the field gradient, for the upper (lower) interferometer. In the case of the maximum achievable magnetic field gradient allowed by our MOT coils during the interferometer sequence ($B_m'^{u} = 12$~G/cm, $B_m'^{l} = 9$~G/cm estimated from the maximum separation $\Delta z=3.9$~cm), the estimated relative shift $\Delta\phi_M$ due to this term in a gradiometer with $T=20$ ms is $\Delta\phi_M \simeq 30$ mrad. 

Experimental tests of this estimation have been conducted by applying a magnetic field gradient $B_m'$ during the interferometer sequence. In particular, by turning on the magnetic field gradient only between the interferometer pulses (and removing the artificial gradient), we observed no appreciable differential phase accumulation between the two interferometers. It is worth mentioning that this observation is consistent with the small phase shift $\Delta\phi_M$ expected, since the sensitivity at small ellipse angles ($\theta< 100$~mrad) degrades to 100~mrad, due to systematic errors in the ellipse fitting for our noise level \cite{Wang2016}. 

\begin{figure*}
    \begin{center}
        \includegraphics[width=0.7  \textwidth]{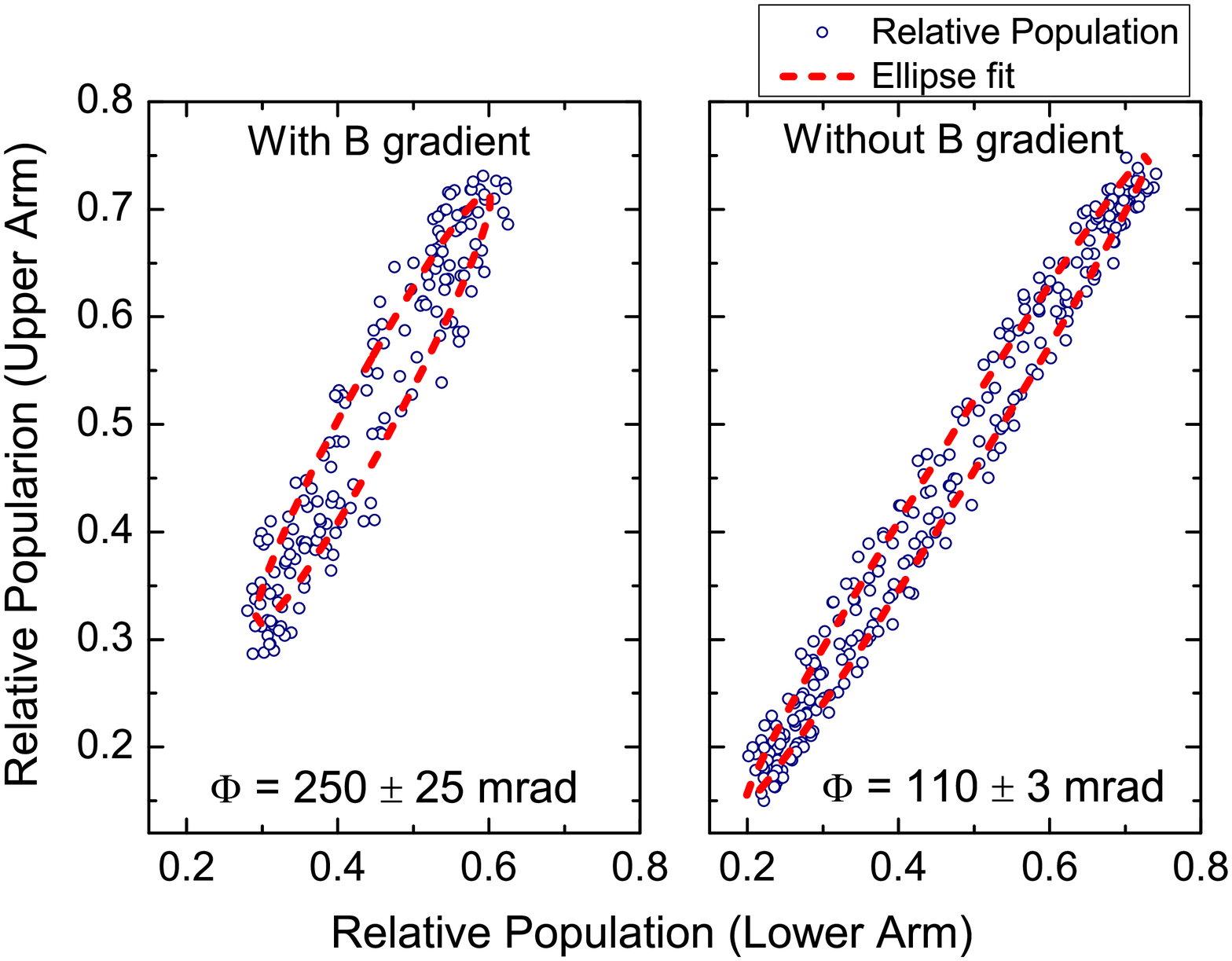}
        \caption{Comparison of fitted ellipses for the case where a magnetic gradient is applied throughout the interferometer (left) and when there is no magnetic gradient (right), where zero phase difference is expected. The presence of a phase difference $\phi=110(3)$ mrad for the case with zero field is a result systematic errors in the ellipse fitting. A study of the fitting errors with artificially generated phase data shows that for phase angles $\phi > 100$ mrad the systematic fitting errors fall in line with the uncertainty of the fit. From this we can see that for an applied magnetic gradient of 12~G/cm, we obtain a gradient induced phase shift of 250(25) mrad (left), due to effect of the magnetically sensitive excited state.} \label{fig.B-Ellipses}
    \end{center}
\end{figure*}	

The situation becomes more complicated when a magnetic field gradient is applied over the \textit{entirety} of the interferometer duration. Here, a further phase contribution arises from the small, but non-zero effect of the upper $^3$P$_1$ magnetically sensitive state, which is coupled to the ground state by the red light during the pulses. Indeed, when a magnetic field gradient $B_m'$ is applied over the whole interferometer sequence, as shown in figure~\ref{fig.B-Ellipses}, a small but non-negligible differential phase $\Phi_{QZ}=(250\pm25)$~mrad has been observed.

Although an exhaustive explanation of this additional effect is not the subject of the present paper, we notice that these measurements demonstrate the expected low sensitivity of a $^{88}$Sr Bragg gradiometer to external field gradients. Indeed, all the experimental tests have been conducted with magnetic field gradients at least $10^3$ times larger than those typically used in similar measurements conducted on alkali atoms \cite{Zhou2010}. As a matter of fact, by applying such a large magnetic field gradient $B_m'$ on a rubidium interferometer, one would expect to observe a huge differential phase shift of $\Phi^{Rb}_{QZ}= 1.4\times10^3$~rad,  completely spoiling the interferometer coherence itself, due to the differential phase shift acquired across a single atomic cloud of typical size \cite{Damico2016}. As a result, the observed differential phase shifts on a $^{88}$Sr gradiometer are about $10^5$ times less than for a gradiometer based on Rb. 
Moreover,  we envision being able to perform a future precision measurement of the magnetic shift effect at the second order in the field amplitude of the ground state in strontium (diamagnetic term) by using a similar measurement configuration, employing an even higher sensitivity red Bragg gradiometer.


\section{Conclusions} 
We reported on the first gradiometer based on Bragg atom interferometry of ultra-cold $^{88}$Sr atoms. Using a high-power laser source at 689~nm, detuned from the narrow intercombination transition, we could both drive the Bragg transitions and efficiently launch two cold atomic clouds from a single MOT.  We are able to obtain a higher interferometer contrast, up to 40\% at interferometer time $T=80$ ms, demonstrating a lower contrast decay rate than previously observed \cite{Mazzoni2015}. We characterize the sensitivity of our gradiometer by introducing an artificial gradient, reaching $1.5 \times 10^{-5}$ ~s$^{-2}$ after 1000~s integration time. Most significantly, the predicted insensitivity to magnetic field gradients of strontium atoms has been demonstrated here for the first time. In particular, the observed low sensitivity, of about $10^{5}$ times less than Rb, allows the operation of the gradiometer even in presence of magnetic field gradients up to 12~G/cm, large enough to prevent other gradiometers based on alkali atoms from working. While the small size of our cell limits the maximum baseline of the interferometer, thus limiting the sensitivity to gravity gradients, the key features of this new interferometer have been shown.  We envision the use of this newly developed gradiometer in future precision measurements of the shift of the ground state of strontium due to the diamagnetic term and future precision measurements of gravitational fields. A strontium Bragg interferometer could also be the basis of future tests of fundamental physics \cite{Tarallo2014} and high accuracy measurements of Newtonian gravitational constant G \cite{Rosi2017G}. 


\section{Acknowledgments} We would like to thank G. Rosi for helpful discussions and C. W. Oates  for a critical review of the manuscript. We would also like to thank W. Bowden for his assistance on improving our data fitting techniques while on secondment at Florence. We acknowledge financial support from INFN and the Italian Ministry of Education, University and Research (MIUR) under the Progetto Premiale ``Interferometro Atomico'' and PRIN-2015. We also acknowledge
support from the European Union's Seventh Framework Programme (FP7/2007-2013 grant agreement 250072 - ``iSense'' project and grant agreement 607493 - ITN ``FACT'' project).

\section*{References}


%

\end{document}